\documentstyle[12pt]{article}
\begin{document}
\baselineskip=24pt
\preprint{IMSc 92/41}
\preprint{hep-lat@ftp/9210024}
\begin{center}
\large{DUAL OF 3-DIMENSIONAL PURE SU(2) LATTICE GAUGE} \\
\vspace{0.2cm}
{THEORY AND THE PONZANO-REGGE MODEL}

\vspace{4cm}

\bf{RAMESH ANISHETTY, SRINATH CHELUVARAJA} \\
and {H.S.SHARATCHANDRA$^{1}$} \\
The Institute of Mathematical Sciences \\
Madras - 600 113, INDIA\\
and\\
\vspace{0.2cm}
\bf{MANU MATHUR}\\
Satyendra Nath Bose National Centre for Basic Sciences \\
DB 17, Sector -1, Salt Lake \\
Calcutta - 700 064, INDIA
\end{center}

\vspace{4cm}

\noindent $^1$e-mail address : ramesha,srinath,sharat@imsc.ernet.in

\vspace{0.2cm}
\noindent PACS No. 11.15.Ha

\newpage

\vspace{4cm}

\noindent{\bf{ABSTRACT}}

\vspace{0.5cm}

By carrying out character expansion and integration over all link variables,
the partition function of 3-dimensional pure SU(2) lattice gauge theory is
rewritten in terms of 6j symbols.  The result is Ponzano-Regge model of
3-dimensional gravity with a term that explicitly breaks general coordinate
invariance.  Conversely, we show that dual of Ponzano-Regge model is an SU(2)
lattice gauge theory where all plaquette variables are constrained to the
identity matrix and therefore the model needs no further regularization.  Our
techniques are applicable to other models with non-abelian symmetries in any
dimension and provide duality transform for the partition function.

\newpage

Duality transformation [1] for statistical mechanical and quantum field
theoretic models have time and again proved to be a powerful tool for
understanding the models.  For instance, it has provided an understanding of
quark confinement in 3 - and 4- dimensional U(1) lattice gauge theories.
Until recently these techniques were available only for models with abelian
global or local symmetries.  Recently some of us [2] have developed the
techniques for non-abelian lattice gauge theories in the Hamiltonian
formalism.  Perhaps it is more useful to have the technique directly for the
partition function, where it is easier to extract the implications and to do
numerical calculations.  Here we show how this can be done using pure SU(2)
lattice gauge theory in 3-(Euclidean) dimensions as an example.

\vspace{0.5cm}

Experience has shown that the duality transformation is tied up with
Fourier analysis of the partition function, and the dual theory is naturally
related to the dual lattice.  It is no different now.  To be more specific, the
steps in the transformation are : (i) Carry out harmonic analysis (in this
case the character expansion) on the Boltzmann factor.  (ii) Isolate the
dependence on, and carry out the integration over, each statistical
variable.  (iii) Interpret the resulting constraints on the dual variables.
Further steps depend on the kind of information that is being sought.  For
instance, one may solve the constraints on the dual variables.

\vspace{0.5cm}

Early in the literature of lattice gauge theory, step(1) was carried out [3],
with a view to get an efficient strong coupling expansion.  However, it was
felt that steps (ii) and (iii) are intractable in the non-abelian case.  We
demonstrate here that the constraints can be handled and the resulting theory
has an elegant  geometrical interpretation.

\vspace{0.5cm}

As mentioned earlier, in this letter we consider only pure SU(2) lattice
gauge theory in 3-Euclidean dimensions.  We find that the partition function
can be rewritten as a sum over products of 6j symbols.  In this case, we
have the added bonus that the result is almost identical to the model of
3-dimensional gravity proposed by Ponzano and Regge [4].  There is an extra
term due to which the partition function is no longer invariant under a
refinement of the simplicial decomposition.  Conversely, Ponzano-Regge model
in the dual form is an SU(2) lattice gauge theory, in which every plaquette
variable is constrained to the identity matrix.

\vspace{0.5cm}

Consider the partition function of 3-dimensional pure SU(2) lattice gauge
theory.

$$
Z \,=\, \prod_{ni}\, \int \,dU (ni)\, {\prod_{\stackrel{nij}{i < j}}}
\quad exp \,{\frac{\beta}{4}}\, (tr U(nij) + h.c-4 )      \eqno(1)
$$

Here an SU(2) matrix U(ni) is associated with every link ni of a
3-dimensional cubic lattice.  The action involves plaquette variables

$$
U(nij) \quad = \quad U(ni)\, U(n+i,j)\, U^{+}(n+j,i)\, U^{+}(nj) \eqno(2)
$$

The integration over configurations is by using the Haar measure dU.
$\beta\, >$ 0 is the coupling constant.

As the action involves only the trace of SU(2) matrices, we have the character
expansion,

$$
exp \quad {\frac{\beta}{4}}\, (tr U+h.c.-4)\quad=\quad \sum_{J}\,
C_{J} {(\beta)} \chi^J (U)  \eqno(3)
$$

\noindent where the summation is over all non-negative half-integral values
of J.  $\chi^J$(U) stands for the trace in the (irreducible representation)
labelled by J of the SU(2) group element U.  If the eigenvalues of U are exp
$i{\frac{\omega}{2}}$ then,

$$
\chi^J(U) \quad = \quad {\frac{sin (2J+1) \omega/2}{sin \omega/2}}  \eqno(4)
$$

The coefficients $C_J {(\beta)}$ of the character expansion are related to the
Bessel functions [3],

$$
\begin{array}{rcl}
C_J {(\beta)} & =  & {\frac{2(2J+1)e^{-\beta}}{\beta}}\quad I_{2J+1}{(\beta)}\\
              & \\
              & \simeq & (2J+1)\, exp - {\frac{1}{2\beta}} J (J+1)-\beta
\end{array} \eqno(5)
$$

Here the last term gives the asymptotic form for large J.

Therefore Z (eqn.1) may be written as,

$$
Z \,= \, \prod_{ni}\, \int  dU(ni)\, \prod_{\stackrel{nij}{i < j}} \quad
\sum_{J(nij)}\quad C_{J(nij)} {(\beta)} \chi^{J(nij)}\quad (U(nij))  \eqno(6)
$$

In order to integrate over U(ni)'s we have to extract the dependence of each
$\chi^{J} (U(nij))$ on U(ni).  This can be done using

$$
\chi^{J} (UV) \quad = \quad D^J_{MN} (U) D^{J}_{NM} (V)     \eqno(7)
$$

\noindent where D$^{J}_{MN}$ (U) are the Wigner D-functions (generalized to
the SU(2) case).  Eqn.(7) is a consequence of the property that the matrix
representing the group element UV is simply the product of matrices for the
individual elements U and V.

Using eqn(7), Z (eqn.6) is expressed in terms of D$^J_{MN}$ (U)'s for
all the link variables.  For each link variable, there are four D$^J$'s
coming from the four plaquettes incident on the link.  In order to
integrate over the link variables, we first combine D$^J$'s pairwise,
using the identity :

$$
D^{J_{1}}_{M_{1}N_{1}} (U)\, D^{J_{2}}_{M_{2}N_{2}} (U) \quad = \quad
\sum_{J_{3}M_{3}N_{3}}\, (2J_{3} + 1)\left( \begin{array}{lll}
 J_{1} & J_{2} & J_{3}\\

 M_{1} & M_{2} & M_{3}   \end{array}  \right)
$$
$$
\times \quad D^{J_{3}}_{M_{3}N_{3}} (U)^{\ast} \left( \begin{array}{lll}
 J_{1} & J_{2} & J_{3} \\

 N_{1} & N_{2} & N_{3}     \end{array}  \right)		\eqno(8)
$$

\noindent where the ranges of M$_{3}$, N$_{3}$ and J$_{3}$ are governed by
the 3jm symbols present. (8) is the Clebsch-Gordan decomposition of the direct
product of two IR's J$_{1}$ and J$_{2}$.  For any choice of orientations,
it is found necessary to also combine $D^J$(U) with a $D^{J} (U)^\ast$.
This can be done using,

$$
D^J_{MN}\,(U)^\ast \quad = \quad (-1)^{M-N}\, D^J_{-M-N} (U) \eqno(9)
$$

Thus the product of four D$^{J}$'s for each link is reduced to sums involving
two D$^{J}$'s.  Now the integration over link variables can be completed
using,

$$
\int dU \quad D^{J_{1}}_{M_{1}N_{1}}\,(U) \quad D^{J_{2}}_{M_{2}N_{2}} (U)^\ast
\quad = \quad \frac{1}{2J_{1} + 1} \, \delta_{J_{1}J_{2}}\,
\delta_{M_{1}M_{2}}\, \delta_{N_{1}N_{2}}    \eqno(10)
$$

This way, integration over all link variables can be carried out.  This
results in an expression for the partition function involving sums over
J's, M's and N's of products of 3jm symbols.  It turns out that the summations
over M's and N's can be explicitly carried out to express the partition
function in terms of 6j symbols. (The underlying reason for this
is local gauge invariance which requires that the 3jm symbols at each site
appear in an invariant combination).  For the calculation, the phase and
other factors have to be carefully kept track of.  Two of us [5] have
developed a modified form of diagrammatic representation [6] of the angular
momenta to carry out these calculations unambiguously.  The result can
be given a geometrical interpretation as follows.

It is natural to associate J(nij) (eqn.6) with the links dual to plaquette
(nij).  These dual links form a cubic lattice.  Thus the four D$^{J}$ (U)'s
for each link are associated with the four links of the plaquette dual to
U.  These dual links are paired (to use eqn.8) in the following specific
way.  The vertices of the dual lattice are alternately labelled
`even' and `odd'.  For each dual plaquette, the dual links meeting at an
`odd' vertex are paired.  The summation index J$_{3}$ resulting on
applying eqn.(8 ) for the associated D$^{J}$'s is represented on the
diagonal which with these dual links completes a triangle i.e. the
diagonal which joins the `even' vertices.  This representation is consistent
because the same J$_{3}$ results from both pairings of this dual plaquette,
a consequence of $\delta_{J_{1}J_{2}}$ in eqn.(10).  Also, this is
suggestive of the triangle rule satisfied by $J_{1}, J_{2}$ and $J_{3}$
(eqn.(8)).  This prescription of drawing face diagonals on the dual cubes
divides each dual cube into five tetrahedrons of which the central
tetrahedron is made entirely of the diagonals.

We are now in a position to express the result of integrating over all
link variables in a simple way :

$$
Z \,=\,   \sum_{\{J(\tilde{l}),J(d)\}} \quad \left( \prod_{\tilde{l}}\,
C_{J{(\tilde{l})}} {(\beta)} \right) \quad
\left( \prod_{J(d)} \,(2J(d) +1) \right)
$$
$$
\times \quad \prod_{{t}}\quad ( 6j symbol)_{t}  \eqno(11)
$$

Here $\tilde{l}$, d and t stand respectively for the dual links, face
diagonals joining the `even' dual vertices and the tetrahedrons thereby
obtained on the dual lattice.  (6j symbol)$_{t}$ is the 6j symbol
represented by the tetrahedron t  in the standard notation [4].

In this form, 3-dimensional pure SU(2) lattice gauge theory has close
resemblance with the Ponzano-Regge model of 3-dimensional gravity which has

$$
Z_{PR}\,=\, \sum_{J(l)}\quad \left( \prod_{l} (-1)^{2J(l)}\, (2J(l)+1)
\right)\quad
\left( \prod_{t} e^{-i\pi \Sigma_{i} J_{i} (t)} \, (6j symbol)_{t} \right)
\eqno(12)
$$

The major difference is that the weight factor for some of the edges (i.e.
for the dual links) is $C_{J}{(\beta)}$ in lattice gauge theory instead of
(2J+1).  This has a drastic consequence.  The partition function (12) has
the remarkable invariance [7] under an arbitrary Alexander transformation
of the simplices.  This invariance is absent in (11).  The other (rather
minor) difference is that the incidence number of the tetrahedrons on an
edge is fixed to be four in case of lattice gauge theory.  This is a
consequence of the cubic lattice and the plaquette action we started with.
(Owing to this all phase factors of (11) and (12) match).  This constraint
is absent in the Ponzano-Regge model.

These considerations can be quantified by noting that for large $\beta$

$$
C_{J} (\beta) \sim (constant)\, (2J+1)\quad \frac{1}{\beta^{3/2}}
\left( 1 + 0 (\frac{1}{\beta}) \right).      \eqno(13)
$$

\noindent Thus with a suitable normalization of the action, we get,
$C_{J} (\beta) \rightarrow (2J+1)$ as $\beta \rightarrow \infty$ and we
recover the Ponzano-Regge action.  In other words,

$$
{\ell}im_{\vec{\beta} \rightarrow \infty}\, \beta^{3/2}\,
exp - \beta(2-tr U) \,=\,
(const.) \delta (U-1)      \eqno(14)
$$

This means, the Ponzano-Regge action may be restated as a SU(2) lattice
gauge theory in which each plaquette variable is constrained to the unit
matrix.  Therefore only global degrees of freedom remain [8] and the
partition function depends only on the topology of the space-time manifold.
Thus the duality transformation provides a direct proof of the results
obtained in [8] using invariance under the Alexander transformations.  It also
shows that the Ponzano-Regge model stands well defined as it is, and there
is no need to regularize it further.

These results can also be understood from a naive analysis of the continuum
theories,[9].  The partition function of 3-dimensional pure SU(2) (continuum)
gauge theory may be rewritten as,

$$
Z\,=\, \int d{\vec{E_{i}}}(x) d{\vec{A_{i}}}(x)\quad exp\, \int d^{3} x
\left( -\, {g^2}\, {\vec{E^{2}_{i}}} (x) + i \epsilon_{ijk}
\vec{E}(x)_k (\partial_{i} \vec{A}_{j} - \partial_{j} \vec{A_{i}} +
\vec{A_{i}} \times \vec{A_{j}} ) (x) \right)        \eqno(15)
$$

\noindent by introducing an auxiliary field ${ E^{a}_{i} (x) }$
where i = 1,2,3 labels the space-time index and a = 1,2,3 labels the SU(2)
index.  g is the gauge coupling constant.  If we formally set $g$\ =\ 0,
 we recover 3-dimensional gravity with A$^{a}_{i}$ and E$^{a}_{i}$
playing the roles of the connection and the dreibein respectively.  The
${\vec{E}^{2}_{i}}$ term has internal Lorentz invariance but explicitly breaks
general coordinate invariance.  Being an ultralocal term, it may be formally
absorbed in the measure.  Thus one may regard 3-dimensional SU(2) gauge
theory as gravity quantized with a measure that explicitly breaks general
coordinate invariance.

In this letter, we have developed techniques for obtaining the dual form of
partition functions with non-abelian global or local symmetries.  In case
of 3-dimensional pure SU(2) lattice gauge theory, we have obtained an
additional bonus by establishing its close relation with the Ponzano-Regge
model.  Further we have shown that the dual form of Ponzano-Regge model
provides a simple explanation of its properties.

\newpage

\noindent {\bf{REFERENCES}}

\begin{enumerate}

\item \underline{Phase Transitions and Critical Phenomena} Vol.1, ed.
C. Domb and M.H. Green (Academic Press, N.Y. 1972). \\
J.B. Kogut, Rev. Mod. Phys. {\bf{51}} 659 (1979) . \\
Robert Savit, Rev. Mod. Phys. {\bf{52}} 453 (1980).
\item Ramesh Anishetty and H.S. Sharatchandra, Phys. Rev. Lett. {\bf{65}}
813 (1990). \\
B. Gnanapragasam and H.S. Sharatchandra, Phys. Rev. {\bf{D 45}} R 1010
(1992).\\
Ramesh Anishetty, G.H. Gadiyar, Manu Mathur and H.S. Sharatchandra,
Phys. Lett. {\bf{B 271}} 391 (1991).
\item J.M. Drouffe and J.B. Zuber, Phys. Repts. {\bf{102}} 1 (1983).
\item G. Ponzano and T. Regge, in \underline{Spectroscopic and Group
Theoretic Methods in Physics}, ed. by F. Bloch (North-Holland, Amsterdam,
1968).
\item Srinath Cheluvaraja and H.S. Sharatchandra, Dual Form of the
Partition Function of Pure SU(2) Lattice Gauge Theory in 3-(Euclidean)
Dimensions, imsc-Th-92/42
\item See, for example, D.A. Varshalovich, A.N. Moskalev, and V.K.
Khersonskii,\\
\underline{Quantum Theory of Angular Momentum} (World Scientific, 1988).
\item V.G. Turaev and O.Y. Viro, "State sum invariants of 3-manifolds and
quantum 6j-symbols",  LOM 1 report 1990 (to be published).
\item H. Ooguri and N. Sasakura, Mod. Phys. Lett. A{\bf{6}} 3591 (1991).\\
Mizoguchi and Tada, Phys. Rev. Lett. {\bf{68}} 1795 (1992).
\item F.A. Lunev, "Three dimensional Yang-Mills theory in gauge invariant
variables", Moscow State University Preprint (1992).
\end{enumerate}

\end{document}